\documentclass[12pt]{article}

\usepackage{graphicx}

\begin{document}

\begin{titlepage}



\centerline{\large \bf { Non-thermal leptogenesis and baryon asymmetry}}
\centerline{\large \bf {in different neutrino mass models}}

\vskip 1cm

\centerline{G. Panotopoulos}

\vskip 1cm

\centerline{Department of Physics, University of Crete,}

\vskip 0.2 cm

\centerline{Heraklion, Crete, GREECE}

\vskip 0.2 cm

\centerline{email:{\it panotop@physics.uoc.gr}}
\begin{abstract}
In the present work we study non-thermal leptogenesis and baryon asymmetry in the universe in different
neutrino mass models discussed recently. For each model we obtain a formula relating the reheating
temperature after inflation to the inflaton mass. It is shown that all but four cases are excluded
and that in the cases which survive the inflaton mass and the reheating temperature after inflation
are bounded from below and from above.
\end{abstract}

\end{titlepage}

\section{Introduction}

The Standard Model (SM) of particle physics~\cite{Langacker:1995hi}
is a very successful theoretical framework for all low-energy phenomena.
However, it is widely considered to be a low-energy limit of some underline
fundamental theory. Perhaps the most direct evidence for physics beyond
the SM is the recent discovery that neutrinos have small but finite
masses~\cite{Fukuda:1998mi, Ahmad:2002jz, Apollonio:1999ae}. A simple
and natural way to explain the tiny neutrino masses is via the seesaw
mechanism~\cite{seesaw}. According to that, the existence of super-heavy
right-handed neutrinos is postulated and the smallness of the masses of the
usual SM neutrinos is due to the largeness of the masses of the new neutrinos.
Solar, atmospheric, reactor and accelerator neutrino experiments (for a summary
of three-flavour neutrino oscillation parameters see e.g.~\cite{Maltoni:2004ei})
seem to indicate neutrino masses in the sub-eV range ($0.001~eV < m_{\nu} < 0.1~eV$),
which implies that heavy right-handed neutrinos weigh
$\sim 10^{10}~GeV-10^{15}~GeV$~\cite{Ellis:2004hy}.

On the other hand, the baryon asymmetry in the universe (BAU) is one of the most challenging problems
for modern cosmology. Both Big-Bang Nucleosynthesis~\cite{Steigman:2005ys} and CMB data (for example
from WMAP~\cite{Bennett:2003bz}) show that in the universe one baryon corresponds approximately to one
billion photons. This very small number should be computable in the framework of the theory of the
elementary particles and their interactions we know today. Nowadays, the most popular way to obtain
the BAU is through leptogenesis~\cite{Fukugita:1986hr}. Initially a lepton asymmetry is generated
through the out-of-equilibrium decays of right-handed neutrinos and then the lepton asymmetry is partially
converted to baryon asymmetry through the non-perturbative ``sphaleron'' effects~\cite{Kuzmin:1985mm}.
In general leptogenesis can be thermal or non-thermal. Thermal leptogenesis usually requires very high
reheating temperature after inflation~\cite{Buchmuller:2002rq}. This can be problematic because of the
gravitino constraint. In supersymmetric models (for reviews in supersymmetry see e.g.~\cite{Nilles:1983ge}
and for supersymmetry in cosmology see e.g.~\cite{Feng:2003zu}) with spontaneous supersymmetry breaking the
superpartner of the graviton, the gravitino, gets a mass depending on how the supersymmetry is broken. In
gravity mediated supersymmetry breaking the gravitino mass is in the range $m_{3/2}=100~GeV-1~TeV$ and the
gravitino (if not the lightest supersymmetris particle) is unstable with a lifetime larger than Nucleosynthesis
time $t_{N} \sim 1~sec$ and dangerous for cosmology. This gravitino problem~\cite{Ellis:1982yb} can be avoided
provided that the reheating temperature after inflation is bounded from above in a certain way,
namely $T_R \leq (10^6-10^7)~GeV$~\cite{Cyburt:2002uv}.

Therefore one can see that heavy right-handed neutrinos can have important implications
both for particle physics and cosmology. Various neutrino mass models~\cite{Altarelli:2002hx, NimaiSingh:2004pi}
have been proposed and their predictions on neutrino masses and mixings
have been studied thoroughly. The requirement for the right baryon asymmetry in the universe as well as for the
right phenomenology for light neutrino masses and mixings puts severe constraints on right-handed neutrinos.
Recently six concrete neutrino mass models were discussed and a comparison of numerical predictions on baryon
asymmetry for these models was presented~\cite{Sarma:2006zk}. Two of the models were almost consistent with the
observed BAU, while the rest of them predicted either a small ($\eta \leq 10^{-19}$) or a large
($\eta \geq 10^{-6}$) baryon asymmetry. The analysis was performed in the framework of thermal leptogenesis.
The aim of the present work is to study the same models  in the framework of non-thermal leptogenesis and derive
the constaints on the inflaton mass and the reheating temperature after inflation.

Our work is organized as follows. After this introduction we review the six neutrino mass models in section 2
and we discuss non-thermal leptogenesis for these models in the third section. Our results are presented in
section 4 and we conclude in the last section.

\section{Review of the different neutrino mass models}

Here we give a brief review of the six neutrino mass models~\cite{NimaiSingh:2004pi} discussed recently
in~\cite{Sarma:2006zk}. The interested reader can find more details in~\cite{NimaiSingh:2004pi, Sarma:2006zk}.
In particular, all the information about the models are collected in Appendix A of~\cite{Sarma:2006zk}. There
is one normal hierarchical model (NHT3), two inverted hierchical  models (InvT2A, InvT2B) and three degenerate
models (DegT1A, DegT1B, DegT1C). According to seesaw mechanism, the light left-handed neutrino mass
matrix $m_\nu$, the heavy right-handed neutrino mass matrix $M_R$ and the Dirac neutrino mass matrix $m_D$ are
related as follows
\begin{equation}
m_\nu = m_D M_R^{-1} m_D^T
\end{equation}
where $M_R^{-1}$ is the inverse of $M_R$ and $m_D^T$ is the transpose of $m_D$.
The predicted values of the neutrino mass-squared differences and mixing parameters are shown in table 1.

\small
\begin{table}[tbp]
\begin{center}
\begin{tabular}{cccccc}
\hline
Type&$\Delta m^{2}_{21}[10^{-5}eV^{2}]$&$\Delta m^{2}_{23}[10^{-3}eV^{2}]$&$\tan^{2}\theta_{12}$&$\sin^{2}2\theta_{23}$&$\sin\theta_{13}$\\
\hline
DegT1A&$8.80$&$2.83$&$0.98$&$1.0$&$0.0$\\
DegT1B&$7.91$&$2.50$&$0.27$&$1.0$&$0.0$\\
DegT1C&$7.91$&$2.50$&$0.27$&$1.0$&$0.0$\\
InvT2A&$8.36$&$2.50$&$0.44$&$1.0$&$0.0$\\
InvT2B&$9.30$&$2.50$&$0.98$&$1.0$&$0.0$\\
NHT3&$9.04$&$3.01$&$0.55$&$0.98$&$0.074$\\
\hline
\end{tabular}
\hfil
\caption{\footnotesize
 Predicted values of the  solar and atmospheric neutrino mass-squared
differences  and three mixing parameters (from~\cite{Sarma:2006zk}).}
\end{center}
\end{table}


\small
\begin{table}[tbp]
\begin{center}
\begin{tabular}{cc|c}
\hline
Type   &  Case (i):~$|M_{j}|$            &  Case (ii):~$|M_{j}|$\\
\hline
DegT1A
& $4.28\times 10^9$, $1.16\times10^{10}$,
& $3.47\times10^7$, $9.42\times10^7$, \\
& $3.84\times10^{13}$
& $3.81\times10^{13}$\\
DegT1B
& $4.05\times10^7$, $6.16\times10^{11}$,
& $3.28\times10^5$, $4.98\times10^9$,\\
& $7.6\times10^{13}$
& $7.6\times10^{13}$\\
DegT1C
& $4.05\times10^7$, $6.69\times10^{12}$,
& $3.28\times10^5$, $4.85\times10^{11}$,\\
& $6.99\times10^{12}$
& $7.81\times10^{11}$\\
InvT2A
& $3.28\times10^{8}$, $9.70\times10^{12}$,
& $2.64\times10^6$, $7.92\times10^{10}$,\\
& $6.79\times 10^{16}$
& $6.70\times10^{16}$\\
InvT2B
& $5.6527\times10^{10}$, $5.6532\times10^{10}$,
& $4.5971\times10^8$ ,$4.5974\times10^8$,\\
& $5.38\times10^{16}$
& $5.34\times10^{16}$\\
NHT3
& $6.51\times10^{10}$, $7.97\times10^{11}$,
& $5.27\times10^8$,$6.45\times10^9$,\\
& $1.01\times10^{15}$
& $1.01\times10^{15}$\\
\hline
\end{tabular}
\hfil
\caption{\footnotesize
 The three right-handed Majorana neutrino masses in GeV (from~\cite{Sarma:2006zk}).}
\end{center}
\end{table}


\small
\begin{table}[tbp]
\begin{center}
\begin{tabular}{ccc|cc}
\hline
Type  &  Case (i):~$\epsilon$   & Case (ii):~$\epsilon$  & Case (i):~$\eta$   & Case (ii):~$\eta$ \\
\hline
DegT1A & $2.10\times10^{-6}$    & $1.71\times10^{-8}$  & $4.99\times10^{-9}$ & $4.06\times10^{-11}$ \\
DegT1B & $2.66\times10^{-18}$   & $2.16\times10^{-20}$ & $1.60\times10^{-23}$ & $1.30\times10^{-25}$ \\
DegT1C & $1.74\times10^{-18}$   & $1.69\times10^{-20}$ & $1.05\times10^{-23}$ & $1.02\times10^{-25}$ \\
InvT2A & $1.59\times10^{-14}$   & $1.27\times10^{-16}$ & $9.94\times10^{-19}$ & $7.96\times10^{-21}$ \\
InvT2B & $1.47\times10^{-2}$    & $1.62\times10^{-4}$  & $5.40\times10^{-5}$  & $5.94\times10^{-7}$  \\
NHT3   & $5.90\times10^{-7}$    & $4.78\times10^{-9}$  & $2.17\times10^{-9}$  & $1.76\times10^{-11}$ \\
\hline
\end{tabular}
\hfil
\caption{\footnotesize
 Calculation of $CP$ asymmetry $\epsilon$  and baryon asymmetry $\eta$ for each neutrino mass model
(from~\cite{Sarma:2006zk}).}
\end{center}
\end{table}


In thermal leptogenesis for the SM case the BAU $\eta \equiv n_B/n_\gamma = 6.1 \times 10^{-10}$ is computed by
the formula~\cite{Sarma:2006zk}
\begin{equation}
\eta = 0.0216 \kappa \epsilon
\end{equation}
where the $CP$ asymmetry $\epsilon$ is defined as
\begin{equation}
\epsilon = \frac{\Gamma - \bar{\Gamma}}{\Gamma + \bar{\Gamma}}
\end{equation}
with $\Gamma=\Gamma(N_1 \rightarrow l_L \phi^{\dag})$ and $\bar{\Gamma}=\Gamma(N_1 \rightarrow \bar{l_L} \phi)$
the decay rates, while the dilution factor $\kappa$ is determined by numerical integration of Boltzmann
equations. However it can be estimated by~\cite{Sarma:2006zk}
\begin{equation}
\kappa = \frac{1}{2 \sqrt{9+K^2}}
\end{equation}
for $0 \leq K \leq 10$
and by
\begin{equation}
\kappa = \frac{0.3}{K (lnK)^{0.6}}
\end{equation}
for $10 \leq K \leq 10^6$, with $K$ the decay parameter $K=\tilde{m}_1/m^*$, where $m^*$ is the equilibrium
neutrino mass $m^* = 1.08 \times 10^{-3}~eV$ and $\tilde{m}_1$ is the effective neutrino mass defined as
\begin{equation}
\tilde{m}_1 = \frac{v^2 (h h^{\dag})_{11}}{M_1}
\end{equation}
with $v$ the electroweak scale, $M_1$ the mass of $N_1$ and $h$ the matrix for the neutrino Yukawa
couplings. The three right-handed neutrino masses for each model are shown in table 2 while the $CP$ asymmetry
and baryon asymmetry are shown in table 3. The Dirac neutrino mass matrix $m_D$ can be either the charged
lepton mass matrix $m_l$ (case (i)) or the up-quark mass matrix $m_{u}$ (case (ii)). We see that NHT3 and
DegT1A models are almost consistent with the observed BAU, while the rest of the models lead either to very
small baryon asymmetry, $\eta \leq 10^{-19}$ (DegT1B, DegT1C, InvT2A), or to large baryon
asymmetry, $\eta \geq 10^{-6}$ (InvT2B).

\section{Non-thermal leptogenesis}

We start by introducing three heavy right-handed neutrinos (one for each family) $N_i, i=1,2,3$ with
masses $M_1, M_2, M_3$, which interact only with leptons and Higgs through Yukawa couplings. In supersymmetric
models the superpotential that describes their
interactions with leptons and Higgs is~\cite{Hamaguchi:2001gw}
\begin{equation} \label{decay}
W_1=Y_{ia} N_{i} L_{a} H_{u}
\end{equation}
where $Y_{ia}$ is the matrix for the Yukawa couplings, $H_u$ is the superfield of the Higgs doublet that
couples to up-type quarks and $L_a$ ($a=e,\mu,\tau$) is the superfield of the lepton doublets. Furthermore,
we assume that after the slow-roll phase of inflation, the inflaton decays dominantly to right-handed neutrinos
through Yukawa couplings and for supersymmetric models the interaction is described by the
superpotential~\cite{Fukuyama:2005us}
\begin{equation}
W_2=\sum_{i} \: \lambda_{i} S N_{i}^{c} N_{i}^{c}
\end{equation}
where $\lambda_{i}$ are the couplings for this type of interaction and
$S$ is a gauge singlet chiral superfield for the inflaton. With such a superpotential the inflaton decay
rate $\Gamma_\phi$ is given by~\cite{Fukuyama:2005us}
\begin{equation}
\Gamma_\phi \equiv \Gamma(\phi \rightarrow N_{i} N_{i})=\frac{1}{4 \pi} |\lambda_{i}|^2 M_{I}
\end{equation}
The reheating temperature after inflation $T_R$ is given by~\cite{Lazarides:2001zd}
\begin{equation}
T_R=\left ( \frac{45}{4 \pi^3 g_{*}} \right )^{1/4} \: ( \Gamma_\phi~M_{pl} )^{1/2}
\end{equation}
where $M_{pl}$ is Planck mass and $g_{*}$ is the effective number of relativistic degrees of freedom at the
reheating temperature. For the reheating temperatures that we shall consider all the particles are relativistic
and for MSSM $g_{*}=915/4=228.75$, while for SM $g_{*}=427/4=106.75$.

Any lepton asymmetry $Y_{L} \equiv n_{L}/s$ produced before the electroweak phase transition is partially
converted into a baryon asymmetry $Y_{B} \equiv n_{B}/s$ via sphaleron effects~\cite{Kuzmin:1985mm}.
The resulting $Y_B$ is
\begin{equation}
Y_{B}=C \: Y_{L}
\end{equation}
with the fraction $C$ computed to be $C=-8/15$ in the MSSM and $C=-28/79$ in the SM~\cite{Harvey:1990qw}. The
lepton asymmetry, in turn, is generated by the $CP$-violating  out-of-equilibrium decays of the heavy neutrinos
\begin{equation}
N \rightarrow l H_{u}^{*}, \quad N \rightarrow \bar{l} H_{u}
\end{equation}
In the framework of non-thermal leptogenesis the lepton asymmetry is given
by~\cite{Asaka:1999yd, Fukuyama:2005us}
\begin{equation}
Y_L=\frac{3}{2} \: BR(\phi \rightarrow N_1 N_1) \: \frac{T_R}{M_{I}} \: \epsilon
\end{equation}
where $M_I$ is the inflaton mass, $T_R$ the reheating temperature after inflation, $\epsilon$ the $CP$ asymmetry
and BR is the branching ratio for the decay of the inflaton to the lightest heavy right-handed neutrino. The
decay is kinematically allowed provided that
\begin{equation}
M_{I} > 2M_{1}
\end{equation}
We will assume that $BR \approx 1$, that is the inflaton decays practically only to the lightest of the
right-handed neutrinos. This is possible even if the inflaton is heavy enough to decay to all right-handed
neutrinos as long as $|\lambda_1|^2 \gg |\lambda_2|^2, |\lambda_3|^2$.
Combining the above formulae we obtain
\begin{equation} \label{7}
Y_{B}=C \: Y_L = C \: \frac{3}{2} \: \frac{T_R}{M_{I}} \: \epsilon
\end{equation}
or
\begin{equation}
T_R=\left ( \frac{2 Y_B}{3 C \epsilon} \right )~M_I
\end{equation}
From the WMAP data~\cite{Bennett:2003bz} we know that
\begin{equation}
\eta_{B} \equiv \frac{n_{B}}{n_{\gamma}}=6.1 \times 10^{-10}
\end{equation}
If we recall that the entropy density for relativistic degrees of freedom is $s=h_{eff} \frac{2 \pi^2}{45} T^3$
and that the number density for photons is $n_{\gamma}=\frac{2 \zeta(3)}{\pi^2} T^3$, one easily obtains for
today that $s=7.04 n_{\gamma}$. Thus for $Y_B$ we have
\begin{equation}
Y_B = 8.7 \times 10^{-11}
\end{equation}
Following~\cite{Asaka:1999yd} we shall consider that $M_1 \geq 100 \: T_R$, because in that case the
neutrino $N_1$ is always out of thermal equilibrium. Finally we recall that $M_I > 2 M_1$.

\section{Results}

Now we can present our results. We shall begin with the SM case first and we shall use for the fraction $C$ the
SM value, namely $C=-28/79$. For each neutrino model (12 cases in total) the $CP$ asymmetry $\epsilon$ as well
as the right-handed neutrino mass $M_1$ are known. Therefore we have i) a formula relating the reheating
temperature to the inflaton mass, ii) a lower bound for the inflaton mass $M_I > 2 M_1$, and iii)
an upper bound for the reheating temperature $T_R \leq 0.01~M_1$. Furthermore, using the relationship
between $T_R$ and $M_I$ we are able to convert the upper limit for $T_R$ to a corresponding upper limit
for $M_I$ and also the lower limit for $M_I$ to a corresponding lower limit for $T_R$. So both $T_R$ and $M_I$
are bounded both from above and from below. Let $T_R^{min}$ and $T_R^{max}$ be the lower and higher value for
the reheating temperature respectively. Then $T_R^{min} < T_R \leq T_R^{max}$ and obviously it is required
that $T_R^{max} > T_R^{min}$, which is not satisfied for all cases. In fact most of the cases are excluded.
The only cases for which the constraint is satisfied are: \\
- DegT1A, case (i), for which:
\begin{equation}
8.56 \times 10^9~GeV < M_I \leq 5.49 \times 10^{11}~GeV
\end{equation}
\begin{equation}
6.67 \times 10^5~GeV < T_R \leq 4.28 \times 10^7~GeV
\end{equation}
- NHT3, case (i), for which:
\begin{equation}
1.3 \times 10^{11}~GeV < M_I \leq 2.35 \times 10^{12}~GeV
\end{equation}
\begin{equation}
3.6 \times 10^7~GeV < T_R \leq 6.51 \times 10^8~GeV
\end{equation}
-InvT2B, case (i), for which:
\begin{equation}
1.13 \times 10^{11}~GeV < M_I \leq 5.09 \times 10^{16}~GeV
\end{equation}
\begin{equation}
1.25 \times 10^3~GeV < T_R \leq 5.65 \times 10^8~GeV
\end{equation}
-InvT2B, case (ii), for which:
\begin{equation}
9.2 \times 10^8~GeV < M_I \leq 4.55 \times 10^{12}~GeV
\end{equation}
\begin{equation}
9.29 \times 10^2~GeV < T_R \leq 4.6 \times 10^6~GeV
\end{equation}
One can see from the results presented above that inflationary models in which $M_I \sim 10^{13}~GeV$, like
e.g. chaotic~\cite{Linde:1983gd} or natural~\cite{Freese:1990rb} inflation, are compatible only with one
neutrino model (InvT2B, case (i)). Furthermore, for a concrete inflationary model with a given inflaton mass
our results allow us to know what the reheating temperature must be and also what the inflaton decay
rate $\Gamma_\phi$ is and what the inflaton Yukawa coupling $|\lambda_1|$ is. For example, in chaotic or
natural inflation we obtain
\begin{eqnarray}
M_I & \sim & 10^{13}~GeV \\
T_R & \sim & 10^5~GeV \\
\Gamma_\phi & \sim & 10^{-8}~GeV \\
|\lambda_1| & \sim & 10^{-10}
\end{eqnarray}
At this point we should add a comment regarding the gravitino constraint. If we add supersymmetry in order to
address the gravitino problem, then the expression for the baryon asymmetry in MSSM will change slightly by a
numerical factor of order one. So we could use the results obtained so far for the SM case. If we require
that $T_R \leq (10^6-10^7)~GeV$ then we see that the models InvT2B and DegT1A are already compatible with the
gravitino constraint, the model NHT3 is marginally compatible (for the lower values for $T_R$) with the
gravitino constraint and finally the model InvT2B can be made compatible with the gravitino constraint lowering
the upper bound for $T_R$
\begin{equation}
1.25 \times 10^3~GeV < T_R \leq (10^6-10^7)~GeV
\end{equation}

\section{Conclusions}

In the present work we have studied non-thermal leptogenesis in six neutrino mass models proposed earlier and
discussed recently in the literature. For each model we have obtained a formula relating the inflaton
mass $M_I$ to the reheating temperature after inflation $T_R$. In fact according to this formula $T_R$ is
proportional to $M_I$. Hence, the bigger the inflaton mass the bigger the reheating temperature. In a concrete
inflationary model (chaotic~\cite{Linde:1983gd}, natural~\cite{Freese:1990rb}, supersymmetric
hybrid~\cite{Copeland:1994vg} etc) with a given mass for the inflaton, the right baryon asymmetry implies
a certain reheating temperatute after inflation. This in turn implies a certain decay rate for the inflaton
field and a certain value for the inflaton Yukawa coupling. Furthermore, kinematical reasons and the
requirement for non-thermal leptogenesis lead to a lower and an upper bound both for $M_I$ and $T_R$.
Our results show that in most of the neutrino models under study the lower bound is not compatible with
the upper bound and therefore only four cases survive. If we also take into account the gravitino
constraint $T_R \leq (10^6-10^7)~GeV$, then in one of these cases the reheating temperature is even more
constrained.


\section*{Acknowlegements}

We would like to thank T.~N.~Tomaras for a critical reading of the manuscript. This work was supported by the EU grant MRTN-CT-2004-512194.

\end{document}